# Signature of high temperature superconductivity in electron doped Sr$_2$IrO$_4$

Y. J. Yan[1], M. Q. Ren[1], H. C. Xu[1], B. P. Xie[1, 2], R. Tao[1], H. Y. Choi[3], N. Lee[3], Y. J. Choi[3], T. Zhang[1, 2], D. L. Feng[1, 2*]

1 State Key Laboratory of Surface Physics, Department of Physics, and Advanced Materials Laboratory, Fudan University, Shanghai 200433, China
2 Collaborative Innovation Center of Advanced Microstructures, Fudan University, Shanghai 200433, China
3 Department of Physics and IPAP, Yonsei University, Seoul 120-749, Korea

**Sr$_2$IrO$_4$ was predicted to be a high temperature superconductor upon electron doping since it highly resembles the cuprates in crystal structure, electronic structure and magnetic coupling constants. Here we report a scanning tunneling microscopy/spectroscopy (STM/STS) study of Sr$_2$IrO$_4$ with surface electron doping by depositing potassium (K) atoms. At the 0.5-0.7 monolayer (ML) K coverage, we observed a sharp, V-shaped gap with about 95% loss of density of state (DOS) at E$_F$ and visible coherence peaks. The gap magnitude is 25-30 meV for 0.5-0.6 ML K coverage and it closes around 50 K. These behaviors exhibit clear signature of superconductivity. Furthermore, we found that with increased electron doping, the system gradually evolves from an insulating state to a normal metallic state, via a pseudogap-like state and possible superconducting state. Our data suggest possible high temperature superconductivity in electron doped Sr$_2$IrO$_4$, and its remarkable analogy to the cuprates.**

The search of high temperature superconductors (HTSC) has long been the pursuit of condensed matter physics [1]. Recently, 5$d$ transition metal oxide Sr$_2$IrO$_4$ has attracted much attention, because it possesses several distinct characteristics that are believed to be important for the high temperature superconductivity [2-16]. Sr$_2$IrO$_4$ is isostructural to La$_2$CuO$_4$, which adopts the same quasi-two-dimensional (2D) layered perovskite structure of K$_2$NiF$_4$ type [2-4]. The IrO$_2$ layers form a square lattice of Ir$^{4+}$ ions with a nominal 5d$^5$ configuration, and there is effectively one hole per Ir$^{4+}$. Due to the cooperation between spin-orbit coupling (SOC), crystal field and Coulomb interaction with comparable strength, a pseudo-spin $j$=1/2 ($j$ being the total angular momentum) antiferromagnetic (AFM) Mott insulating state is realized in Sr$_2$IrO$_4$ at low temperature [5,6]. The low-energy magnetic excitations can be described by a $j$=1/2 AFM Heisenberg model [5-11], and the nearest-neighbor AFM exchange interactions $J$ is about 60 - 100 meV, which is comparable to that of cuprates [9,10].

The remarkable resemblance between $Sr_2IrO_4$ and cuprates makes $Sr_2IrO_4$ a good candidate for exploring unconventional HTSC upon carrier doping. Indeed *d*-wave superconductivity by electron doping was predicted by several theoretical studies [11-13]. Meanwhile a triplet *p*-wave pairing state in the hole-doped regime was also suggested when the Hund's coupling is comparable to SOC [13]. Experimentally, electron doping was realized in $Sr_2IrO_4$ by La substitution, oxygen deficiency or surface K dosing [4,14-16], while hole doping was realized by Rh substitution of Ir [17]. Unique electronic state with nodal quasiparticles and an antinodal pseudogap was found in the electron-underdoped regime by angle resolved photoemission spectroscopy (ARPES) [14-16], resembling the underdoped cuprates. However, no experimental evidence of superconductivity has been found up to now. The search of superconductivity in doped $Sr_2IrO_4$ remains of great interest.

In this article, we report a low temperature STM study on electron-doped $Sr_2IrO_4$ via *in situ* surface K dosing [14,18]. $Sr_2IrO_4$ single crystals were grown by a flux method using $SrCl_2$ flux. The sample was mounted onto the holder by conductive epoxy. After pre-cooled at 77 K in vacuum ($<1\times10^{-10}$ *torr*), the samples were cleaved and then immediately transferred into the STM head stabilized at 4.5 K. Au stripes were evaporated onto the surfaces of the sample and the holder by a raster-like mask to enhance tunneling channels. Thereafter, potassium atoms were evaporated onto the whole surface of the sample using commercial SAES alkali metal dispensers. Samples were kept at about 80 K during these operations. The surface coverage was precisely controlled by K flux and growth time. For STM measurements, Pt tip was used for all the measurements after being treated on Au (111) surface. STM topography is taken in the constant current mode, and dI/dV spectrum is collected using a standard lock-in technique with modulation frequency f = 975 Hz.

Pristine $Sr_2IrO_4$ was cleaved in the vacuum at 77K, leaving charge-balanced SrO-terminated surface [19]. Figure 1(b) shows a 50 × 50 $nm^2$ topographic image measured at 77 K. The atomically resolved image with a square lattice is shown in the top inset. The Fourier transform (FT) of Fig. 1(b) (bottom inset) shows two sets of spots. $q_1$ is the Bragg spot of SrO lattice, while $q_2$ is from a $\sqrt{2}$ R45º reconstruction, which could be due to the rotation of the $IrO_6$ octahedra, as reported in ref. [2]. Four types of defects are observed in the cleaved surface, as indicated by the arrows. The dI/dV spectrum measured at 77K, away from defects exhibits an insulating energy gap of about 700 meV, which is consistent with the previous STM report [19]. We found that the type 3 defect, which is likely oxygen defects [19], drastically suppresses the insulating gap as shown in Fig. 1(c). While other types of defects do not affect the local density of state (LDOS) significantly.

At temperatures lower than 30 K, we found that the tunneling cannot be obtained on pristine $Sr_2IrO_4$ due to drastically increased sample resistance (even upon surface K dosing). Thus to introduce conducting channels, we evaporated Au contacts onto part of the cleaved surface through masks (Fig. 1(a) and Fig. s1). Then the tunneling to the region close to Au contacts is achieved at low temperature. Fig. 1(d) shows a topographic image of such region (taken at 20 K), some scattered Au clusters with high LDOS can be seen on the surface. A decreased insulating gap is observed in this region, indicating enhanced conductivity (Fig. 1(e), red curve). This might be caused by some small amount of electrons transferred from the gold clusters to $Sr_2IrO_4$. Meanwhile the insulating state is still maintained on $Sr_2IrO_4$ far away from the Au contacts. Thereafter, K atoms were evaporated on the surface to dope electron carriers [14,18], resulting the final sample configuration sketched in Fig. 1(a). After depositing certain amount of K (0.5~1ML coverage), metallic state can be observed in the region close to the Au contacts (Fig. 1(e), blue curve) [20]. Fig. 1(f) is a typical topographic image of such region with 0.6 ML K coverage, the K atoms form clusters with a typical size of several nanometers. We then measured tunneling spectrum in this area, as shown in Fig. 1(g). One found that for the K coverage of 0.5~0.7ML, there is a sharp, V-shaped gap structure in dI/dV spectrum, which is symmetric with respect to $E_F$. The gap magnitude changes with K coverage (25-30 meV for 0.5 ML and 0.6 ML, 10 meV for 0.7 ML). For all the observed gaps, about 95% of DOS vanishes near $E_F$. In the literature, a gap like structure at $E_F$ may have various causes, such as density wave (DW), some pseudogap, and superconductivity. For the first two cases, as observed in most previous STM studies, the gaps are usually not fully opened and give large residual DOS at $E_F$ [21]. Only for a superconducting state, a fully gapped dI/dV spectrum with coherence peaks is usually observed, and a V-shaped gap appears in the presence of nodes. The sharpness and nearly fully vanished DOS at $E_F$ of the observed gap and the presence of coherent peaks make pseudogap state or DW state unlikely the origin. In addition, no sign of charge density modulation is observed here and in the previous photoemission measurements [14]. Therefore, the observed gap is most likely caused by superconductivity.

To visualize the spatial distribution of this symmetric gap, we show the dI/dV map taken close to the gap edge at the bias voltage ($V_b$) of 20 meV in Fig. 2(a) (measured at 20 K), taking sample with 0.6 ML K dosing as an example (see also Fig. s2 for dI/dV maps at other energies). The spatial variation of tunneling conductance is reflected by the false color. There is a strong spatial inhomogeneity, or phase separation between several types of patches, as highlighted by the color and representative dotted boundaries. The corresponding representative dI/dV spectra are shown in Fig. 2(c). With increasing conductance, there are roughly four types of regions:

    1. The low conductance region (the dark blue region) is the insulating region, which is characterized by the insulating gap larger than 100 meV in the corresponding spectrum (Fig. 2(c), curve 1);

    2. The transition region (the light blue and light green region), where the insulating gap is much smaller in the center of this region, and gradually filled up at the boundary (Fig. 2(c), curves 2 and 3);

    3. The pseudogap-like region (the green and yellow region), where the superconducting gap at 25-30 meV appears, but the dominating feature is a

pseudogap-like feature at around the positive bias of 60 meV (defined as E*), as shown by Fig. 2(c), curve 4;

4. The possible superconducting region (the red region), where a small symmetric V-shape gap-like structure ubiquitously exists, as shown by curve 5 in Fig. 2(c). The gap is about 25-30 meV. This region is more homogeneous than others, and is the majority one. The coherence peak is stronger here, while the pseudogap-like feature is weak or absent.

In Fig. 2(b), we show the dI/dV map taken around the pseudogap energy at $V_b$ = 70 meV. This map is clearly anti-correlated with Fig. 2(a), which may suggest that the pseudogap-like state competes with the possible superconductivity here. Because such a competition was found for the cuprate superconductor Bi2212 as well [22], it is also a demonstration of the possible superconducting nature of the observed gap near $E_F$. Similar electronic inhomogeneity has been observed in cuprates as well, which is attributed to inhomogeneous carrier distribution [21,23]. Since the K atoms were evaporated onto the sample surface holding at 80 K, which results in clustered surface morphology (Fig. 1(e)), the carrier concentration here is likely to have spatial inhomogeneity. As shown in Fig. 1(b), $Sr_2IrO_4$ is an insulator with an energy gap of about 700 meV. The observed lineshape evolution and transition between different regions clearly indicate that: 1. Mott gap of $Sr_2IrO_4$ is continuously suppressed by the increasing electron doping, and eventually disappears; 2. a pseudogap-like structure develops, which presumably is the antinodal pseudogap observed previously by ARPES [14]; 3. with further electron doping, the system evolves into a possible superconducting state. Such behaviors share strong similarities with those of cuprates, in which a Mott insulating state evolves into a pseudogap state and then superconducting state with increasing hole doping [21].

The inhomogeneity of LDOS distribution is weakened with increasing K-coverage (See Fig. s2 and Fig. s3). However the pseudogap-like feature and superconducting gap feature can be observed in a wide range of K-coverage, as shown in Figs. 3(a) and 3(b). The energy scale of E* decreases gradually with the increasing K coverage as indicated by the arrows in Fig. 3(a). The distance between the two coherence peaks of the typical dI/dV spectra defines 2Δ (Δ being the possible superconducting gap). In Fig. 3(b), it is 54 meV for 0.5 ML, 59 meV for 0.6 ML, and decreases to 22 meV for 0.7 ML, respectively. Considering the inhomogeneity of the gap magnitudes, the average Δ is about 28 meV for both 0.5 and 0.6 ML, and 10 meV for 0.7 ML, which are determined from the spatially averaged dI/dV spectra (See Fig. s4). For $Sr_2IrO_4$ with 1 ML K, the typical dI/dV spectrum (taken at 4.5 K) shows an overall flat DOS with only a small dip at $E_F$, indicating a normal metallic phase. As reported in ref. [14] and observed in our ARPES results shown in Fig. s5 of the supplementary materials, $Sr_2IrO_4$ with 1 ML K overlayer shows quantum well states with intense spectral weight, indicating the formation of a metallic state in the K overlayer. Thus the dI/dV spectrum on $Sr_2IrO_4$ with 1 ML K will be contributed by both of the K overlayer and the top layer of $Sr_2IrO_4$. Then the small dip at $E_F$ is possibly a signature of superconductivity in $Sr_2IrO_4$ with 1ML K. With further increasing K coverage, a normal metallic state is observed down to 4.5K in dI/dV spectrum without any signature of superconductivity, which possibly means that the system has been over-doped beyond the superconducting regime, as observed in the cuprates.

To obtain possible superconducting transition temperature ($T_c$) of the K-doped $Sr_2IrO_4$, we measured temperature dependence of the spatially averaged dI/dV spectra with large gap depth and visible coherence peaks, as shown in Figs. 4(a)-(c). For K-coverage of 0.5 - 0.7 ML, the gap gradually fills up and the coherence peaks get weakened as temperature increases. However, even after the coherence peaks disappear (>60K in Fig. 4(a)), there is still a broad V-shaped background. This feature may be also induced by the pseudogap-like state nearby which could coexist with superconducting state. To determine the onset temperature of the gap more quantitatively, we have defined gap depth = 1-ZBC/CCP (ZBC: zero bias conductance, CCP: averaged conductance of two coherence peaks), which is plotted in Fig. 4(d). The gap depth starts to increase rapidly upon cooling at 20 K for 0.7 ML, and at about $50\pm5$ K for both 0.5 ML and 0.6 ML. We assigned these characteristic temperatures as their corresponding $T_c$'s. The ratios of $2\Delta/k_BT_c$ are then estimated to be in the range of 12.5-13.4 for 0.5-0.7 ML K coverages. Such an unusually large $2\Delta/k_BT_c$ ratio has been observed in underdoped cuprates as well [21,23-24].

In Figure 5, we plot $E^*$, gap magnitude $\Delta$ and $T_c$ as a function of the K coverage. $\Delta$ and $T_c$ scales with each other, and appear to saturate at low K coverage, while $E^*$ and the pseudogap measured by ARPES both increase with decreased coverage [14]. These behaviors largely resemble those of the hole doped cuprates [21]. The electron doping of $Sr_2IrO_4$ was already suggested to be the analogue of hole doping of cuprates by early theoretical and experimental reports [11,14], which is hereby further demonstrated by our STM results. A detailed comparison between K doped $Sr_2IrO_4$ and a prototypical cuprate is listed in Table. 1. Furthermore, we have observed many other characteristic features here that were observed in cuprates, for example, the nanometer scale electronic inhomogeneity, the systematic evolution of the electronic states with carrier doping, the pseudogap-like phase and its coexistence and competition with the possible superconducting phase.

Table I: Comparison of K dosed $Sr_2IrO_4$ and hole doped cuprates.

|  | K doped $Sr_2IrO_4$ | hole doped cuprate (Bi2212) |
|---|---|---|
| Electronic states with different carrier concentration | Mott insulator<br>pseudogap-like state<br>possible superconductivity<br>normal metallic state | Mott insulator<br>pseudogap<br>superconductivity<br>normal metallic state |
| Magnetism | pseudo-spin<br>*j=1/2*<br>nearest-neighbor AFM | spin<br>*s=1/2*<br>nearest-neighbor AFM |

|  | exchange interaction $J \sim 0.06\text{-}0.1$ eV | exchange interaction $J \sim 0.12$ eV |
|---|---|---|
| Superconducting gap | V-shape, d-wave? | V-shape, d-wave |
|  | 25-30meV at maximal $T_c$ | ~35 meV at optimal doping |
| Maximal $T_c$ (where gap depth stops decreasing on warming) | 50±5K | ~ 90K |
| $2\Delta/K_B T_c$ | 12.5-13.4 | 7-14 |

In summary, we have systematically studied the electronic states of $Sr_2IrO_4$ with different surface K coverage. At the K coverage of 0.5~0.7 ML, we observed sharp, V-shaped gap with visible coherence peaks, providing strong evidence of possible superconductivity in electron doped $Sr_2IrO_4$. We also demonstrated that with increased surface K coverage, the electronic state of $Sr_2IrO_4$ evolves from an insulating state to a normal metallic state with more than 1 ML K, via a pseudogap-like state and a possible superconducting state sequentially. Our results present a discovery of possible unconventional high temperature superconductivity in *5d* transition metal compounds. Because of the strong SOC in $Sr_2IrO_4$, spin is not a good quantum number, this system thus is the first realization of HTSC whose Cooper pairs are formed by two *j*=1/2 pseudo-spins. The remarkable analogy between this system and hole doped cuprates, particularly the scaling between their Δ's, $T_c$'s, and *J*'s, and the evolution of various electronic states with doping would help to build a universal theme of high temperature superconductivity and solve this long standing mystery.

**Acknowledgements**
We thank Prof. Changyoung Kim for helping set up the collaboration, and Prof. Fa Wang for helpful discussions. This work is supported by the National Science Foundation of China, and National Basic Research Program of China (973 Program) under the grant No. 2012CB921402. The work at Yonsei was supported by the NRF Grant (NRF-2013R1A1A2058155 and NRF-2014S1A2A2028481) and partially by the Yonsei University Future-leading Research Initiative of 2014 (2014-22-0123).

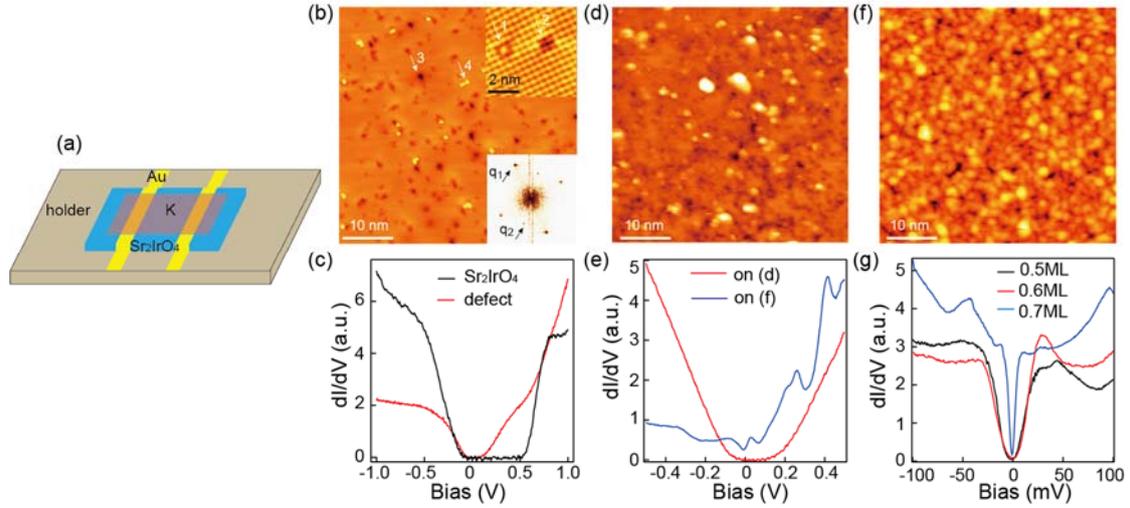

FIG. 1. Surface topography and dI/dV spectrum of $Sr_2IrO_4$ with and without K coverage. (a) Sketch of the sample configuration. (b) Typical topographic image on the SrO-terminated surface of pristine $Sr_2IrO_4$, measured at 77 K. The top and bottom insets represent the atomically resolved image and the FT of data in panel (b), respectively. Four different kinds of defects are indicated by arrows. (c) Spatially averaged dI/dV spectrum on the SrO-terminated surface. An insulating energy gap as large as 700 meV is observed on defect free region, while a reduced gap is observed on defects 3. (d) Typical topographic image of the SrO-terminated surface adjacent to Au contacts (measured at 20 K), in which a few Au clusters are observed as the white spots. (e) Spatially averaged dI/dV spectra measured on panels (d) and (f), showing a reduced insulating gap and a metallic state, respectively. (f) Typical topographic image after depositing 0.6 ML K atoms on the SrO-terminated surface that is adjacent to the Au contacts (measured at 20 K). (g) Representative dI/dV spectra on K doped $Sr_2IrO_4$ with various K coverages (10K for 0.5ML, 20K for 0.6 ML, 4.5K for 0.7 ML), showing possible superconducting gaps.

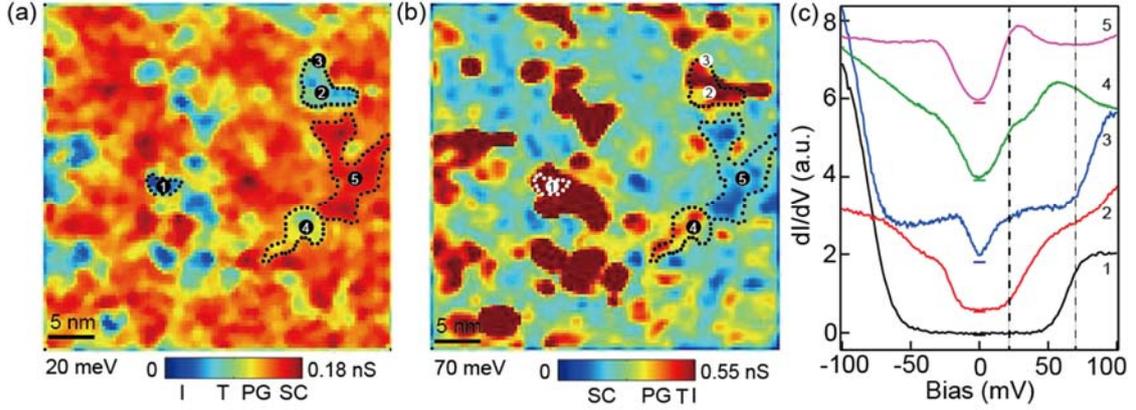

FIG. 2. Gap inhomogeneity of $Sr_2IrO_4$ with 0.6 ML K taken at 20 K. (a)-(b) dI/dV maps taken at $V_b$=20 meV and 70 meV respectively. Representative areas with different electronic states are marked by dotted lines. Each map has 100 × 100 pixels. (c) Typical dI/dV spectra taken at the positions marked by dots in panels (a) and (b), showing the evolution of electronic states across the regions. The horizontal bars indicate zero conductance position of each curve. Dashed lines located at 20 meV and 70 meV are added for guide of the eyes. The labels I, T, PG, and SC on the color bars indicate the insulating, transition, pseudogap-like, and possible superconducting regions, respectively.

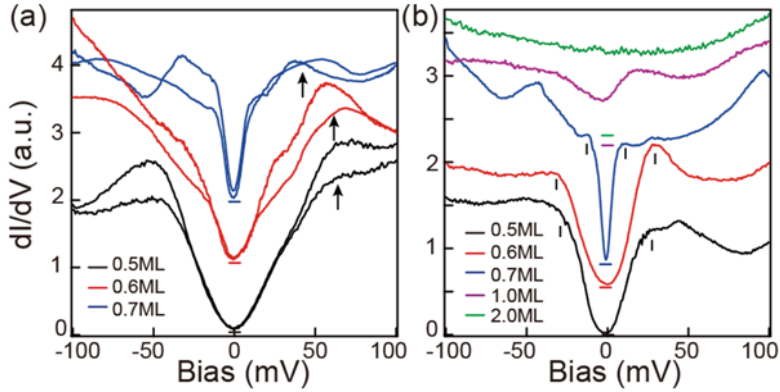

FIG. 3. K coverage dependence of dI/dV spectra. (a) The representative spectra taken at the pseudogap-like regions for various K coverages. Inhomogeneity of the pseudogap-like feature is illustrated by two spectra shown for each coverage. The arrows indicate the averaged energy locations of $E^*$. (b) The representative spectra taken at possible superconducting regions for 0.5-0.7 ML K coverage and arbitrary regions for 1-2 ML K coverage. The curves are offset vertically for clarity, and the horizontal markers indicate zero conductance position of each curve. The data were taken at 10 K for 0.5 ML, 20 K for 0.6 ML and 4.5 K for 0.7-2 ML.

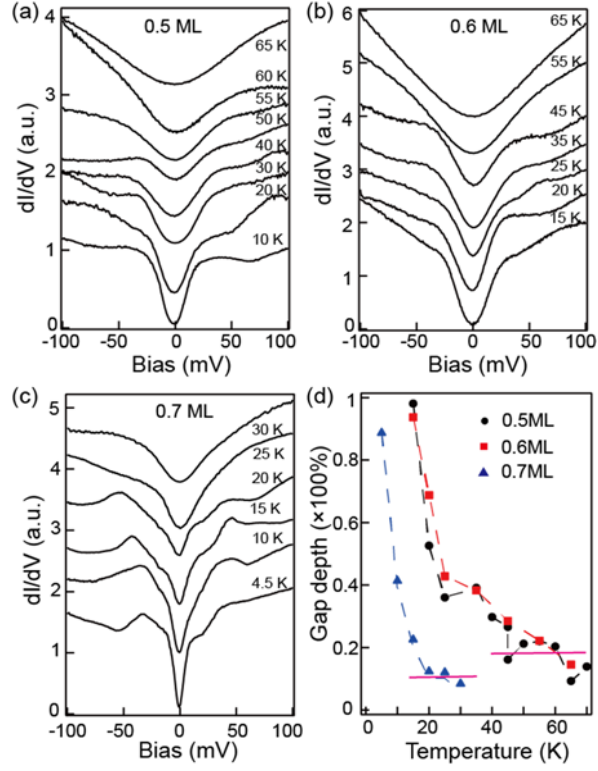

FIG. 4. Temperature dependence of the possible superconducting gap. (a)-(c) Temperature dependence of the spatially averaged dI/dV spectra for K coverage of (a) 0.5 ML, (b) 0.6 ML, and (c) 0.7 ML respectively. The spectra shown here are the average of the dI/dV spectra with large gap depth and visible coherence peaks taken on the SC areas in a 30 × 30 nm$^2$ area. The lineshape variations at different temperature are caused by the lack of precisely tracking of the same location on a clustered surface with strong LDOS inhomogeneity. The curves are offset vertically for clarity. (d) Gap depth (as defined in the main text) as a function of temperature, which decreases gradually with increasing temperature. $T_c$ is defined by the temperature that the gap depth stops decreasing quickly upon warming.

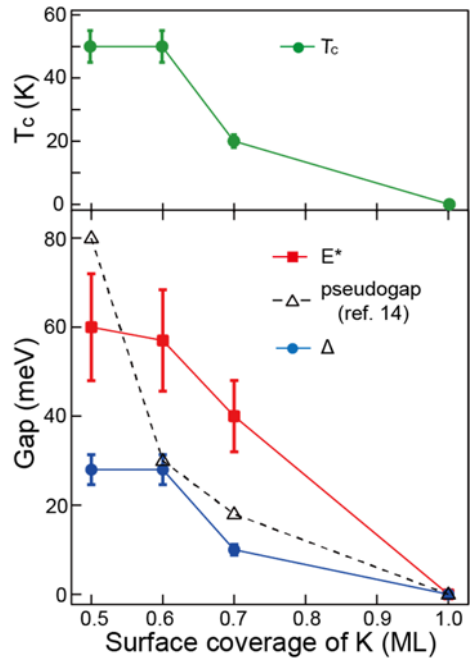

FIG. 5. $E^*$, $\Delta$ and $T_c$ as a function of surface coverage. The gap magnitude of the pseudogap observed at the antinode by ARPES is shown by empty triangles for comparison [14].